# A Key Set Cipher for Wireless Sensor Networks

Subhash Kak

**Abstract**
This paper proposes the use of sets of keys, together with corresponding identifiers, for use in wireless sensor networks (WSN) and similar computing resource-constrained applications. Such a system, with each user assigned a bunch of private key vectors with corresponding public identifiers to generate session keys, is harder to break than where a single key is used. The set of keys and identifiers are generated by a suitable mathematical operation by the trusted party and assigned to users. A specific cryptographic system described in the paper is based on the use of a family of self-inverting matrices derived from the number theoretic Hilbert transform (NHT) in conjunction with the Blom's scheme. In a randomized version of this scheme, the users change their published IDs at will but the parties can still reach agreement on the key by using their individual scaling factors. The random protocol increases the security of the system.

**Introduction**
Wireless sensor networks (WSNs) help to connect a variety of power-constrained devices to the Internet. To ensure security in the communications with these devices a variety of key management schemes (KMS) are used of which the principal ones use key pooling, mathematical algorithms, or public key cryptography. The factors in deciding which scheme to use are based on memory requirements, communication overhead, processing speed, network bootstrapping, connectivity, scalability, extensibility, and energy requirement [1]. Of the mathematical algorithms used, the matrix-based Blom scheme has found many applications such as in High-bandwidth Digital Content Protection (HDCPv1).

Each HDCP-capable device has a unique set of 40 56-bit keys. For each set of values, a special 40-bit public key called a KSV (Key Selection Vector) is created. During authentication, the parties exchange their KSVs using Blom's scheme leading to a shared 56-bit number, which is also used to encrypt data. The HDCP method is used in DVD, HD DVD, Blu-ray Disc players, computer video cards, TVs and digital projectors.

In the basic Blom's scheme [2], the trusted authority gives each user a secret key and a corresponding public identifier to enable any two participants to independently create a shared key for secure communication. The Blom scheme is different from either key pooling or public key techniques in that any two users discover a unique key between the two of them. In contrast, there can be different ways two users can communicate in a network using key pooling, and in public key cryptography, the users can change their private and public keys, if only in principle.

Therefore another way to distinguish between KMS schemes is whether they involve use of single keys or multiple keys between any pair of users.

The use of multiple keys is of special advantage in making the task of the eavesdropper more difficult. Such use can also be of value in forensics. The paradigm of multiple keys for each user has antecedents that go back in history in social systems (in the many ways personhood is defined) [3] and there exist parallels in the biological realm. This idea is of philosophical validity to problems that go beyond safeguarding wireless sensor networks. In digital networks, many individuals maintain different identifiers ranging from two (one at work and another at home) to several. Some of these identities are motivated by the need for anonymity.

In this paper we look at ways of enhancing Blom's matrix based scheme so that it can support key sets for the users. Such an enhancement poses a greater challenge to the eavesdropper as compared to simply increasing the size of the key. The Blom scheme [2] is based on two different non-square matrices X (n×m) and Y (m×n) whose product mod p (p is a suitably large prime) is the symmetric matrix K (n×n). The matrix X is used to generate the users' keys and the matrix Y is used to generate the corresponding public identifiers. An obvious way to have a large key set is to use a set of indexed X and Y matrices. There exist a variety of ways to determine X and Y [4]. We can assume that these indexed matrices are obtained by means of suitable linear transformations on an initial member. For this propose the use of NHT-circulant matrices which are orthogonal in the same $Z_p$ as Blom's matrix.

Given that the key pairs are based on different matrices, the task of breaking the system has been made more difficult. Since the question of security of the Blom scheme is well analyzed [5],[6],[7], here we only describe the construction associated with the generation of the key set and the corresponding system identifiers. We also present a randomized scheme in which each user publishes a subset of the identifiers and changes them with time. This requires a scheme to normalize the keys generated by the use of different identifiers.

**Matrix Based Keys**

In the Blom scheme, the row *i* of X represents the secret key of User *i* ($X_{i1}, X_{i2}, ..., X_{im}$) and the column *i* of Y represents the public identifier of the User *i* ($Y_{1i}, Y_{2i}, ..., Y_{mi}$). Let Alice be User *i* and Bob be User *j*. The size of the key and the public identifier is *m*.

Alice finds the key by multiplying her private key vector with the identifier (public key) of Bob thus getting $K_{ij}$. Likewise, Bob multiplies his private key with the identifier of Alice thus getting $K_{ji}$. Since the matrix K is symmetric, Alice and Bob obtain the same number which can be used as the raw key:

$$K_{ij} = \sum_{\alpha} X_{i\alpha} Y_{\alpha j} \qquad (1)$$

*Definition.* The value of $K_{ii}$ will be called the scale of the key of User $i$. It will also be represented by $S$.

We will see later that the key scale plays a critical role in the determination of the common key in the randomized version of the algorithm.

*Example 1.* Let $X = \begin{bmatrix} 9 & 9 & 6 \\ 1 & 9 & 2 \\ 1 & 0 & 0 \\ 4 & 5 & 2 \\ 0 & 5 & 0 \end{bmatrix}$ mod 11, and $Y = \begin{bmatrix} 7 & 10 & 5 & 1 & 8 \\ 7 & 2 & 6 & 0 & 3 \\ 7 & 7 & 3 & 4 & 4 \end{bmatrix}$ mod 11 (2)

$K = \begin{bmatrix} 3 & 7 & 7 & 0 & 2 \\ 7 & 9 & 10 & 9 & 10 \\ 7 & 10 & 5 & 1 & 8 \\ 0 & 9 & 1 & 1 & 0 \\ 2 & 10 & 8 & 0 & 4 \end{bmatrix}$ mod 11

If Alice is User 2 and Bob is User 4, then

Alice's secret key: (1, 9, 2)
Alice's public identifier: (10, 2, 7)  (3)

Bob's secret key: (4, 5, 2)
Bob's public identifier: (1, 0, 4).  (4)

Their shared key is found by computing the inner product of the secret key and public identifier vectors. The shared key:

As computed by Alice: $1 \times 1 + 9 \times 0 + 2 \times 4 = 9 \mod 11 = 9$
As computed by Bob: $4 \times 10 + 5 \times 2 + 2 \times 7 = 64 \mod 11 = 9$

The key scales of Alice and Bob are 9 and 1, respectively, and they will be called $S_{Alice}$ and $S_{Bob}$.

**Modified Scheme to Generate Key Sets**

The Trusted Authority can generate a large number of equivalent keys by using appropriate matrix transformations on a known solution (Figure 1). Given the constraint of a fixed m (size of the key), the matrix transformation can either be an outer transformation (n×n) or an inner transformation (m×m).

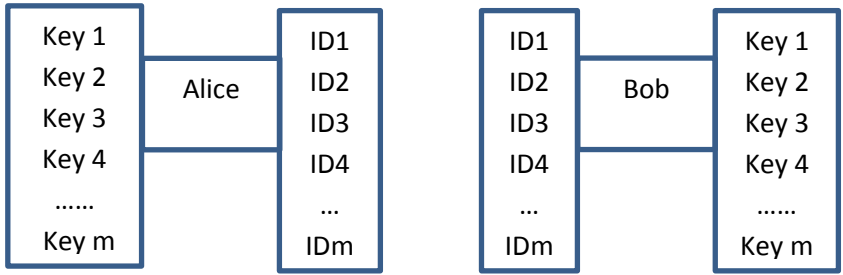

Figure 1. A matrix-based key-set scheme

The keys of User *i* are defined as follows:

Table 1. User *i*'s Key and Identifier Set

| User *i*: Index | Secret Key X(row i) | Public Identifier Y(column i) |
|---|---|---|
| 1 | X(1) | Y(1) |
| 2 | X(2) | Y(2) |
| 3 | X(3) | Y(3) |
| …. | …. | …. |
| M | X(m) | Y(m) |

*Outer transformation*: Form new X and Y by the following transformations:

$$X_{new} = UX \quad \text{and} \quad Y_{new} = YV \tag{5}$$

and

$$K_{new} = UXYV = UKV \tag{6}$$

There are different ways to choose U and V (each of which will be a n×n matrix) but one easy method would be to pick them as symmetric matrices.

*Inner transformation*: Form new X and Y by the following transformations:

$$X_{new} = XR \quad \text{and} \quad Y_{new} = SY \tag{7}$$

so that the product below remains a symmetric matrix:

$$K_{new} = XRSY \tag{8}$$

Specifically, we will consider the following inner transformations:

$$X_{new} = XR \text{ where } Y_{new} = R^T Y, \text{ and } RR^T = wI \text{ mod } p \tag{9}$$

This will change the scales of the keys by the factor w.

The matrix R in (9) represents a set whose transpose is its inverse (up to a constant). It is related to the number theoretic Hilbert transform (NHT) [8], which is derived from the standard discrete Hilbert transform [9]-[11].

The NHT is a circulant matrix with alternating entries of each row being zero and non-zero numbers and transpose modulo a prime is its inverse. The discrete Hilbert transform has applications is a variety of areas of signal processing such as spectral analysis in 2-D reconstruction [12]-[16], multilayered computations [17],[18], and cryptography [19]-[21]. A circulant matrix is used in the mix columns of the Advanced Encryption Standard. Periodic sequences that are orthogonal for all shifts and their potential applications were recently presented [22],[23]. Some further properties of circulant matrices relevant to our paper are given in [24].

*Example 1 Contd*. For Alice as User *i* and Bob as User *j*, let us assume that Alice decided to use Bob's 7th identifier. She sends this information to Bob in the pre-communication handshake so that Bob will use the corresponding secret key. The key found by Alice and Bob would now be:

K = Alice (secret key 7) × Bob (identifier 7) = Bob (secret key 7) × Alice (identifier 7)

If the set of matrices R is so chosen that w in equation (9) equals 1, then the key obtained will be same irrespective of what row in Table 1 is chosen.

**The Matrix R and the Key Set**
Consider the related matrix R, which is identical to NHT matrix with the exception that its 0s have been purged.

$$R = \begin{bmatrix} a_1 & a_2 & a_3 & . & a_N \\ a_N & a_1 & a_2 & a_3 & . \\ . & a_N & a_1 & a_2 & a_3 \\ a_3 & . & a_N & a_1 & a_2 \\ a_2 & a_3 & . & a_N & a_1 \end{bmatrix} \text{ mod } p \tag{10}$$

The sequence $a_1, a_2, a_3, ..., a_N$ may be considered a random sequence. For consideration as a sequence, we assume that the generator is periodic, i.e. $a_{N+i} = a_i$. If we don't insist on normalization, it will satisfy the following properties under the condition:

$$\sum_{i=1}^{N} a_i^2 \text{ mod } p = w \tag{11}$$

$$C(k) = \sum_{i=1}^{N} a_i a_{i+k} \bmod p = 0 \text{ for all } k \neq 0 \tag{12}$$

Property 12 means that the sequence $A = a_1, a_2, a_3, ..., a_N$ may be considered a truly random sequence with autocorrelation function, $C(k)$, that is zero everywhere excepting at k=0. It is also clear that w×A would be another random sequence for which the autocorrelation function at the origin will be $w^2$ mod p and its value everywhere else will be 0. The peak value of the autocorrelation is obtained from the sequence for which $w^2$ mod p = p-1.

Let us go back to Example 1. The matrix required there is 3×3. The general R matrix for which $RR^T=wI$ may be taken to be $\begin{bmatrix} a & b & c \\ c & a & b \\ b & c & a \end{bmatrix}$, but since we want flexibility regarding w, this could be replaced by

$$R = \begin{bmatrix} 1 & a & b \\ b & 1 & a \\ a & b & 1 \end{bmatrix} \tag{13}$$

This leads to the conditions:

$$w = 1 + a^2 + b^2 \bmod p \tag{14}$$

and

$$a + b + ab = 0 \bmod p \tag{15}$$

Equation (15) may be easily solved. We have $b(a+1) = -a \bmod p$. This may be rewritten as

$$b = \frac{p-a}{a+1} \bmod p \tag{16}$$

Since our example was concerning p =11, we can solve (16) for different choices of a and b as given in Table 2 below:

Table 2. Solutions to equation (16) for the 3×3 matrix

| a | 1 | 2 | 3 | 4 | 5 | 6 | 7 | 8 | 9 |
|---|---|---|---|---|---|---|---|---|---|
| b | 5 | 3 | 2 | 8 | 1 | 7 | 6 | 4 | 9 |
| w | 5 | 3 | 3 | 4 | 5 | 9 | 9 | 4 | 5 |

Suppose we pick a=2 and b=3. For this choice the value of w is 3 and therefore the new key will be the old key multiplied by 3. The new X and Y matrices would now be:

$$X_{new}= \begin{bmatrix} 9 & 9 & 6 \\ 1 & 9 & 2 \\ 1 & 0 & 0 \\ 4 & 5 & 2 \\ 0 & 5 & 0 \end{bmatrix} \begin{bmatrix} 1 & 2 & 3 \\ 3 & 1 & 2 \\ 2 & 3 & 1 \end{bmatrix} \mod 11 = \begin{bmatrix} 4 & 1 & 7 \\ 10 & 6 & 1 \\ 1 & 2 & 3 \\ 1 & 8 & 2 \\ 4 & 5 & 10 \end{bmatrix} \qquad (17)$$

Likewise, the new Y matrix will be:

$$Y_{new}=R^T Y_{old} \qquad (18)$$

In other words,

$$Y_{new}= \begin{bmatrix} 1 & 3 & 2 \\ 2 & 1 & 3 \\ 3 & 2 & 1 \end{bmatrix} \begin{bmatrix} 7 & 10 & 5 & 1 & 8 \\ 7 & 2 & 6 & 0 & 3 \\ 7 & 7 & 3 & 4 & 4 \end{bmatrix} \mod 11 = \begin{bmatrix} 9 & 8 & 7 & 9 & 3 \\ 9 & 10 & 3 & 3 & 9 \\ 9 & 8 & 8 & 7 & 1 \end{bmatrix} \qquad (19)$$

The new K matrix is:

$$K_{new} = \begin{bmatrix} 9 & 10 & 10 & 0 & 6 \\ 10 & 5 & 8 & 5 & 8 \\ 10 & 8 & 4 & 3 & 2 \\ 0 & 5 & 3 & 3 & 0 \\ 6 & 8 & 2 & 0 & 1 \end{bmatrix} \qquad (20)$$

Clearly $K_{new} = 3K$ as expected.

It so turns out that the choices for the R matrix for this example (Table 2) belong to two groups:

Group 1:

    1 2 3 ◄
    2 4 6
    3 6 9
    4 8 1 ◄
    5 10 4

6 1 7 ◀
7 3 10
8 5 2
9 7 5
10 9 8

Group 2:

1 1 5 ◀
2 2 10
3 3 4
4 4 9
5 5 3
6 6 8
7 7 2
8 8 7
9 9 1 ◀

In the modified scheme, the secret key and public identifier will be indexed as follows:

Alice's Secret Key and Public Identifier Table (User 2):

Table 3. Alice's key, identifier vectors, and scales

| Index(a,b, c) | Secret Key | Public Identifier | $K_{22} = S_{Alice}$ |
| --- | --- | --- | --- |
| 1 (original) | 1, 9, 2 | 10, 2, 7 | 9 |
| 2 (1, 2, 3) | 10, 6, 1 | 8, 10, 8 | 5 |
| 3 (4, 8, 1) | 7, 2, 4 | 10, 7, 10 | 3 |
| 4 (6, 1, 7) | 5, 3, 6 | 4, 5, 4 | 4 |
| 5 (1, 1, 5) | 4, 9, 5 | 5, 3, 4 | 1 |
| 6 (9, 9, 1) | 3, 4, 1 | 1, 5, 3 | 4 |

Bob's Secret Key and Public Identifier Table (User 4):

Table 4. Bob's key and identifier vectors

| Index(a, b, c) | Secret Key | Public Identifier | $K_{44} = S_{Bob}$ |
| --- | --- | --- | --- |
| 1 (original) | 4, 5, 2 | 1, 0, 4 | 1 |
| 2 (1, 2, 3) | 1, 8, 2 | 9, 3, 7 | 3 |
| 3 (4, 8, 1) | 4, 10, 8 | 3, 1, 6 | 4 |
| 4 (6, 1, 7) | 6, 4, 1 | 10, 7, 9 | 9 |
| 5 (1, 1, 5) | 9, 8, 5 | 5, 10, 9 | 5 |
| 6 (9, 9, 1) | 4, 6, 1 | 1, 2, 4 | 9 |

Many variants of this system may be proposed. This includes the cases where the size of the key itself can vary making the task of the eavesdropper even more complex.

**Use of Random Subset of the Key Set**

We now describe a random variant of the scheme outlined in the previous section where each user randomly publishes a different subset of the public identifier set at different times. This is achieved by the user publishing the random subset and also announcing the time duration for which it is effective.

The challenge is that several matrices are in play in the different public identifiers available to the users and the two parties may not share the same identifier index. This necessitates the use of the scaling factors by the two parties and prior agreement that the scaling be done with respect to a specific index. This agreement can be negotiating in the set-up and it creates an additional parameter adding to the security. Here, in our example, it will be assumed that the scaling will be done with respect to index 1. The key generation will have two steps: in the first one, each user discovers a raw key which may be different for different users; in the second, the raw key is changed using the personal scaling factor by each party.

In our running example, let Alice randomly pick the identifiers 2 and 4, and let Bob randomly pick the identifier 3 and 5 (Figure 2). In other words,

    Alice publishes Index 2, Public ID 8, 10, 8
                 Index 4, Public ID 4, 5, 4

    Bob publishes Index 3, Public ID 3, 1, 6
               Index 5, Public ID 5, 10, 9

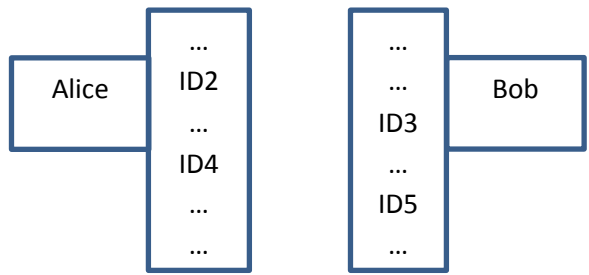

Figure 2. Randomized key-set scheme (each user publishes a subset of the IDs)

Alice computes her scale $K_{22}$ for Bob's Index 3 and Index 5 and likewise Bob computes his scale $K_{44}$ for Alice's Index 2 and Index 4. These scale values are:

    For Alice: Scales for Indices 3 and 5 are: 3 and 1, respectively
    For Bob, Scales for Indices 2 and 4 are: 3 and 9, respectively

In order to initiate the discovery of the common key, Alice picks, say Index 3 ID of Bob, with its sequence 3, 1, 6.

Likewise, Bob picks one of the available IDs of Alice, say Index 4, with the sequence 4, 5, 4.

**Generation of Raw Key**

Alice computes the product (Alice Secret Key 3 × Bob's Identifier 3) = (7, 2, 4) × (3, 1, 6) = 3 mod 11. This is Alice's raw key.

Bob computes the product (Bob's Secret Key 4 × Alice's Identifier 4) = (5, 3, 6) × (4, 5, 4) = 4 mod 11. This Bob's raw key.

**Generation of Final Key**

The final key is computed by normalizing the raw key with respect to the common agreed identifier index value for the communicating parties that has been taken to be 1.

$$\text{Alice's final key} = \text{Alice's raw key} \times \frac{S_{Alice}(CommonIndex)}{S_{Alice}(CurrentIndex)} \qquad (21)$$

$$\text{Bob's final key} = \text{Bob's raw key} \times \frac{S_{Bob}(CommonIndex)}{S_{Bob}(CurrentIndex)} \qquad (22)$$

From Table 3, $S_{Alice}(CommonIndex=1) = 9$ and $S_{Alice}(CurrentIndex=3)=3$, and from Table 4, $S_{Bob}(CommonIndex=1) = 1$ and $S_{Bob}(CurrentIndex=4) = 9$.

Therefore,

Alice's final key= 3×9/3 mod 11 = 9

Bob's final key = 4×1/9 mod 11 = 4 × 5 = 20 mod 11 = 9 \qquad (23)

Suppose the common index chosen was 2, then the values of $S_{Alice}(CommonIndex=2) = 5$ and $S_{Bob}(CommonIndex=2) = 3$ (both from Table 3), and plugging these in, the final key for both the parties turns out to be 5.

The security of the random protocol would depend not only on the size of the key and p, but also on the number of keys in the key set and the manner in which the subsets are chosen and changed in the random version of the scheme.

**Conclusions**

The advantage of the proposed enhancement (both the complete and the randomized versions) to the Blom's scheme is that it increases the cost of breaking the cipher for the eavesdropper. The correct common key can be generated only upon the use of the scaling keys as in equations (21) and (22) and these are not available to the eavesdropper.

The common index used for the normalization of the scaling need not be published by any of the parties and this resides as a secret factor within the node. A description of the use of the cipher proposed in this paper as a formal protocol would require several choices related to the details of the system that will depend on the application in hand.

The Blom-scheme based HDCPv1 was hacked although its use has continued based on legal protections [25] and HDCPv2 is based on a different scheme [26]. The strengthening of the matrix-based Blom's scheme is important and the methods presented in this paper can be an attractive option for resource-limited nodes of a sensor network and in specialized high-bandwidth digital content security.

The Blom scheme is related to threshold secret sharing schemes [27],[28] in the sense that the system is secure unless a certain number of users have been compromised. It will be interesting to explore analogs of space-efficient secret sharing schemes [29],[30] for matrix-based WSN key distribution.

## References

1. C. Alcaraz, J. Lopez, R. Roman, and H. - H. Chen, Selecting key management schemes for WSN applications. Computers & Security 31: 956-966 (2012)
2. R. Blom, An optimal class of symmetric key generation systems, Advances in Cryptology: Eurocrypt 84 (T. Beth, N. Cot and I. Ingemarsson, eds.) LNCS 209, 335-338 (1985)
3. S. Kak, Ritual, masks, and sacrifice. Studies in Humanities and Social Sciences 11, Indian Institute of Advanced Study, Shimla (2004)
4. A. Parakh and S. Kak, Matrix based key agreement algorithms for sensor networks. Proceedings of IEEE Advanced Networks and Telecommunications Conference (ANTS 2011), Bangalore ( 2011)
5. H. Chan, A. Perrig and D. Song, Random key predistribution schemes for sensor networks, IEEE Sumposium on Research in Security and Privacy, 197-213 (2003)
6. C. J. Colbourn and J.H. Dinitz, The CRC Handbook of Combinatorial Designs, CRC Press, 1996)
7. W. Du, J. Deng, Y. S. Han and P. K. Varshney, A pairwise key pre-distribution scheme for wireless sensor networks, Proc. of the 10th ACM conf. on Computer and communications Security, 42-51 (2003)
8. S. Kak, The number theoretic Hilbert transform. arXiv:1308.1688
9. S. Kak, The discrete Hilbert transform. Proc. IEEE 58, 585-586 (1970)
10. S. Kak, Hilbert transformation for discrete data. International Journal of Electronics 34, 177-183 (1973)
11. S. Kak, The discrete finite Hilbert transform. Indian Journal Pure and Applied Mathematics 8, 1385-1390 (1977)
12. S.K. Padala and K.M.M. Prabhu, Systolic arrays for the discrete Hilbert transform. Circuits, Devices and Systems, IEE Proceedings 144, 259-264 (1997)
13. F. W. King, Hilbert Transforms. Cambridge University Press (2009)
14. I.G. Roy, On robust estimation of discrete Hilbert transform of noisy data. Geophysics 78, (2013)
15. S. Kak and N.S. Jayant, Speech encryption using waveform scrambling. Bell System Technical Journal 56, 781-808 (1977)

16. N.S. Jayant and S. Kak, Uniform permutation privacy system, US Patent No. 4,100,374, July 11 (1978)
17. S. Kak, Multilayered array computing. Information Sciences 45, 347-365 (1988)
18. S. Kak, A two-layered mesh array for matrix multiplication. Parallel Computing 6, 383-385 (1988)
19. R. Kandregula, The basic discrete Hilbert transform with an information hiding application. 2009. arXiv:0907.4176
20. V.K. Kotagiri, The 10-point and 12-point number theoretic Hilbert transform. arXiv:1310.3221
21. V.K. Kotagiri, New results on the number theoretic Hilbert transform. arXiv:1310.6924
22. S. Kak, A class of orthogonal sequences. arXiv:1311.6821
23. V.K. Kotagiri, Random residue sequences and the number theoretic Hilbert transform. arXiv:1311.6848
24. S. Kak, Properties of NHT-circulant matrices. http://www.cs.okstate.edu/~subhashk/nhtcirculant_properties.pdf
25. D. Kravets, Intel threatens to sue anyone who uses HDCP crack. Wired (2010)
26. J. Zhao, D. Gu, Y. Li, W. Cheng, On weaknesses of the HDCP authentication and key exchange protocol and its repair. Mathematical and Computer Modelling 55: 19-25 (2012)
27. G.R. Blakley, Safeguarding cryptographic keys. Proceedings of the National Computer Conference 48: 313–317 (1979)
28. A. Shamir, How to share a secret". Communications of the ACM 22 (11): 612–613 (1979)
29. A. Parakh and S. Kak, Online data storage using implicit security. Information Sciences 179: 3323-3331 (2009)
30. A. Parakh and S. Kak, Space efficient secret sharing for implicit data security. Information Sciences 181: 335-341 (2011)